\documentclass[english,preprint,showpacs]{revtex4}
\usepackage{babel}
\usepackage[ansinew]{inputenc}
\usepackage{graphicx}
\usepackage{color}
\usepackage{amsfonts}
\usepackage{amsmath,amssymb}
\usepackage{amsthm,subfigure}
\usepackage{amsbsy}
\usepackage[mathscr]{eucal}

\begin{document}   

\title{On the Manifestly Covariant J\"uttner Distribution and Equipartition Theorem}
\author{Guillermo Chac\'{o}n-Acosta}\email{gca@xanum.uam.mx}
\affiliation{Departamento de F\'\i sica, Universidad Aut\'onoma
Metropolitana-Iztapalapa,  M\'exico D. F. 09340, M\'exico}
\author{Leonardo Dagdug}\email{dll@xanum.uam.mx}
\affiliation{Departamento de F\'\i sica, Universidad Aut\'onoma
Metropolitana-Iztapalapa,  M\'exico D. F. 09340, M\'exico}
\author{Hugo A. Morales-T\'ecotl}\email{hugo@xanum.uam.mx}
\affiliation{Departamento de F\'\i sica, Universidad Aut\'onoma
Metropolitana-Iztapalapa,  M\'exico D. F. 09340, M\'exico}
\affiliation{Instituto de Ciencias Nucleares, Universidad Nacional Aut\'onoma de M\'exico, A. P. 70-543, M\'exico D. F. 04510,  M\'exico}

\begin{abstract}
The relativistic equilibrium velocity distribution plays a key role in describing several high-energy and astrophysical effects.  Recently, computer simulations favored J\"uttner's as the relativistic generalization of Maxwell's distribution for $d=1,2,3$ spatial dimensions and pointed to an invariant temperature.
In this work we argue an invariant temperature naturally follows from manifest covariance. We present a new derivation of the manifestly covariant J\"uttner's distribution and Equipartition Theorem. The standard procedure to get the equilibrium distribution as a solution of the relativistic Boltzmann's equation is here adopted. However, contrary to previous analysis, we use cartesian coordinates in $d$+1 momentum space, with $d$ spatial components. The use of the multiplication theorem of Bessel functions turns crucial to regain the known invariant form of J\"uttner's distribution. Since equilibrium kinetic theory results should agree with thermodynamics in the comoving frame to the gas the covariant pseudo-norm of a vector entering the distribution can be identified with the reciprocal of temperature in such comoving frame. Then by combining the covariant statistical moments of J\"uttner's distribution a novel form of the Equipartition Theorem is advanced which also accommodates the invariant comoving temperature and it contains, as a particular case,  a previous not manifestly covariant form.
\end{abstract}

\maketitle

\section{Introduction}
Incorporating the relativity principles in kinetic theory is crucial not only to understand the theoretical grounds in the description of relativistic many particles systems \cite{stew,dG} but to interpret relativistic high-energy experiments like those involving heavy-ion collisions  \cite{rhic2} as well as phenomena in the astrophysical \cite{chandra} and cosmological realms \cite{wein}. These include, for instance, the use of the  relativistic Bolztmann equation to understand the thermal history of the Universe \cite{peack,bern} and the structure of the Cosmic Microwave Background Radiation spectrum associated to its interaction with hot electrons in galaxy clusters \cite{SZ}.

In the case of equilibrium the history of the relativistic analogue of Maxwell's velocity distribution goes back to F. J\"uttner, who in 1911 turned to relativity to consistently get rid of the contribution of particles with speeds exceeding that of light in vacuum, denoted by $c$, and which are contained in Maxwell's distribution. This was achieved upon replacement of the non-relativistic energy of the free particles by its relativistic form and a maximum entropy principle thus yielding the so called one-particle J\"uttner's distribution \cite{jut}
\begin{equation}\label{MJ}
    {f}= \frac{n }{4\pi kT m^2cK_2(mc^2/kT)}\, e^{-\frac{\sqrt{\mathbf{p}^2c^2+m^2c^4}}{kT}},
\end{equation}
which is written here in momentum rather than velocity space. $m$ is the rest mass of any of the particles forming the gas, $k$ is Boltzmann's constant and $K_2$ is the modified Bessel function of order two \cite{abrwtz}. The remaining  quantities:  particle number density $n$, temperature $T$ and the relativistic spatial  momentum of the particle $\mathbf{p}$, should be here understood as determined by an observer comoving with the gas as a whole. The mass shell condition, $(E/c)^2-\mathbf{p}^2 = m^2c^2$, is assumed along the sequel. In the non-relativistic regime $|\mathbf{p}|\ll mc, mc^2\ll kT$, we have $f\approx {f}_{\mathrm{Maxwell}}$,  where $f_{\mathrm{Maxwell}}$ stands for Maxwell's velocity distribution. Thus particles's speeds beyond $c$ are just an artifact of the non-relativistic approximation.

J\"uttner's distribution in the form (\ref{MJ}) can be disadvantageous in that it is not manifestly covariant, namely it is not expressed in terms of Lorentz tensors which in turn explicitly show its behavior under Lorentz transformations to investigate its description for frames in relative motion. To make it manifestly covariant two key information pieces are needed: the transformation under change of frame of ${f}$ and that of $T$. Both of them have been considered in the literature at large \cite{ei,lan2} and \cite{vankampen1969,neuge,Israel1963,dG,deb}, for temperature and distribution function, respectively.

Upon multiplying (\ref{MJ}) by the characteristic function for the box confining the gas so that $\chi_{_{\mathrm{Box}}}(\mathbf{x})=1$ for $\mathbf{x}$ within the box, and zero otherwise, the resulting distribution $\bar{f}(\mathbf{x},\mathbf{p} ):=\chi_{_{\mathrm{Box}}}(\mathbf{x}){f(\mathbf{p})}$ is defined on the single-particle phase space. An observer would thus determine $\bar{f}$ particles of the gas in a volume $d^3x$ located at $\mathbf{x}$ and having momentum $\mathbf{p}$ within range $d^3p$. Moreover $\bar{f}$ must be Lorentz invariant \cite{Dirac1924,vankampen1969,deb}. This is readily seen in the case of equilibrium \cite{MTW}: for a simple gas the number of particles $\cal N$ is invariant and so ${\cal N}=\int d\mu \bar{f}$, with $d\mu=d^3xd^3p$, must be. Since $d\mu$ is a Lorentz invariant measure due to a compensation between the transformations of the (mass-shell) spatial momentum and space measure, $\bar{f}$ must be so too. Now since $\chi_{_{\mathrm{Box}}}$ is invariant \footnote{Consider two frames $S,S'$ with the gas container lying on $S$. For relative speed $v$, $\gamma=1/\sqrt{1-v^2/c^2}$, Lorentz volume contraction $V'=\gamma^{-1} V$ yields  $\int d^3x'\; {\chi '}_{_{\mathrm{Box}}}(x')= \int d^3x'\; \chi_{_{\mathrm{Box}}}(x), d^3x'=\gamma^{-1}d^3x$ thus proving invariance of $\chi_{_{\mathrm{Box}}}$.} then $f$ is. Hence a manifestly covariant  form of $f$ amounts to manifest invariance and, in particular, it should involve the behavior of temperature under Lorentz transformations. The manifestly invariant J\"uttner distribution was given long ago \cite{sygrg,Israel1963,stew,neuge}. It was determined by introducing spherical coordinates on the pseudosphere associated to the mass shell in relativistic four-momentum space. This provided a rather elegant and fruitful treatment which however does not seem to have been fully appreciated in particular regarding temperature.

Alternatives to (\ref{MJ}) have been proposed recently for which Lorentz covariance is incorporated in a different manner.  In \cite{horwitz1989} a quantum mechanical approach including an invariant  ``historical time" was considered to derive a manifestly covariant Boltzmann equation with stationary solution implying a different ultra-relativistic mean energy-temperature relation whereas in \cite{Lehmann,DunkelHanggi} a maximum relative entropy principle was combined with an invariant group theoretical measure approach to obtain an equilibrium distribution.
Both alternatives have been recently criticized in \cite{debbsch}.
Moreover recent Molecular Dynamics Simulations of a two component one dimensional simple relativistic gas showed agreement with (\ref{MJ}) and temperature defined through an equipartition theorem was shown to be invariant under change of frame \cite{cub}. The study of the two-dimensional case along these lines has been reported in \cite{montakhab} and \cite{sim-eur}. For three spatial dimensions Monte Carlo simulations have been considered favoring also J\"uttner's distribution \cite{peano}. Amusingly, as for temperature, such kind of analysis take us back to the long standing discussion of whether a moving object appears cooler \cite{ei,lan2,cub,montakhab}.

In this work we shall obtain the manifestly invariant J\"uttner distribution by adopting ``cartesian'' rather than spherical coordinates in relativistic ($d$+1)-momentum space \cite{sygrg,Israel1963}. Our approach can be considered as an alternative to such previous derivations of the J\"uttner's distribution  and to others \cite{dG,K&C} in the sense that we avoid adopting any frame along the sequel which otherwise would raise the question on the Lorentz transformation of temperature. We only allude to the frame comoving with the gas at the end of the analysis to relate the kinetic description with thermodynamics in particular to identify temperature with the Lorentz invariant norm of a thermal vector. We once more obtain in this way the thermal four vector introduced long ago \cite{sygrg,Israel1963,stew,neuge} which is formed by the product of the inverse comoving temperature with the four velocity of the gas as a whole. Also the lower dimensional cases recently studied \cite{cub,sim-eur,montakhab,peano} are contained in our results. Hence  comoving temperature is seen to play a role analogue to rest mass of a particle. The compatibility of such an interpretation is further investigated in relation with a manifestly covariant form of the equipartition theorem.

The paper is organized as follows. For the sake of completeness section II briefly reviews the derivation of the J\"uttner distribution as an equilibrium solution of the relativistic Boltzmann equation. This includes a brief description of the two types of approaches: the manifestly covariant one adopting spherical coordinates on the mass shell pseudo-sphere in momentum space and the one adopting the comoving frame and cartesian components. Section III provides the details of our analysis in which we use ``cartesian'' coordinates in momentum space to get the manifestly invariant distribution. In particular the thermal Lorentz vector is here characterized. Its role in the relativistic covariant equipartition theorem is the subject of section IV.  Finally section V summarizes our results. We also describe the behavior of a Planckian distribution when use is made of invariant temperature. The difficulties to define a non-comoving temperature for a gas of massive particles is also touched upon to further stress that invariant comoving temperature seems to be the only consistent possibility to define temperature according to the standard relativistic kinetic theory framework.

Our conventions are the following.  Spacetime components of tensors are denoted by greek indices: $\mu,\nu,\dots=0,1,2,3$, the zero component referring to time whereas spatial components are denoted by latin indices $i,j,\dots=1,2,3$. Einstein sum convention is assumed for both types of indices throughout. The Minkowski metric has components $\eta_{\mu\nu}=\mathrm{diag}\{+,-,-,-\}$.

\section{J\"uttner distribution and the relativistic Boltzmann equation}

Consider a simple relativistic gas composed by identical particles of
mass $m$. To each particle correspond space-time and four-momentum coordinates, respectively given by
$x^{\mu}$ and $p^{\mu}$. The evolution of the
one-particle distribution function of the gas is governed by the
relativistic Boltzmann equation \cite{dG,K&C,G-CGP}
\begin{eqnarray}\label{2.1}
    p^{\mu}{\partial \bar{f}\over \partial x^\mu} +m{\partial
 \bar{f}\mathcal{K}^{\mu}\over \partial p^\mu} &=& {\cal C}[\bar{f}^{*},\bar{f}]\nonumber\\
 {\cal C}[{\bar{f^{*}}},\bar{f}] &:=& \int \left(\bar f^{*}_1 {\bar{f^{*}}}-\bar{f}_1\bar{f}\right)
 \,F\,\sigma\,d\Omega {d^3p_1\over p^0_1}.
\end{eqnarray}
Here ${\cal C}[{\bar{f^{*}}},\bar{f}]$ is the so called collision term and $\bar f^{*}_1\equiv \bar f({\bf x},{\bf p}^{*}_1,t)$, $*$ implying a quantity is evaluated after the collision. $\sigma$ is the differential cross section of the scattering process whereas $\Omega$ is the solid angle. $\mathcal{K}^{\mu}$ denotes an external 4-force while  $F$ is the so called invariant flux, $F=\sqrt{(p^\mu_1p_\mu)^2-m^4c^4}$, and $d^3p_1\over p^0_1$ the invariant measure on mass shell. We shall be interested in the case $\mathcal{K}^{\mu}=0$.

The equilibrium state can be defined so that the entropy production vanishes \cite{Israel1963,neuge,dG,K&C}. In covariant form this means $\frac{\partial S^{\mu}}{\partial x^{\mu}}=0$, with the entropy flux given by
\begin{eqnarray} \label{S3}
S^{\mu} = - kc \int \frac{d^3p}{p^0} p^{\mu}\; \bar{f} \; \mathrm{ln} (h^3 \bar{f}),
\end{eqnarray}
and $h$ is a constant with dimensions to make the argument of the logarithm dimensionless. Zero entropy production requires $\bar{f^{*}} \bar{f^{*}}_1 = \bar{f}\bar{f}_1$ and so the collision term ${\cal C}[\bar{f^{*}},\bar{f}]$ becomes zero too. Such condition can be written as
\begin{equation}\label{2.2}
     \ln{ f^{*}} + \ln{ f^{*}_1 }= \ln{f} + \ln{f_1},
\end{equation}
where $\chi_{_{\mathrm{Box}}}$ canceled out. Eq. (\ref{2.2}) is fulfilled by the \emph{collisional invariants} \cite{K&C} which for the present case become the four-momentum up to a constant (independent of $p^{\mu}$), so that an $f$ solving (\ref{2.2}) takes on the form
\begin{equation}\label{2.3}
    \ln{f} = -\left( \Lambda(x) + \tilde{\Theta}_{\mu}(x)p^{\mu}\right) \; \Leftrightarrow \;
    f = \alpha(x)\exp{\left( -\Theta_{\mu}(x)p^{\mu} \right)}.
\end{equation}
Here the negative sign has been introduced for later convenience, $p^{\mu}$ is the 4-momentum of a particle of the gas and $\alpha=e^{-\Lambda}$. The equilibrium distribution function is fully obtained by requiring consistency between (\ref{2.3}) and the left hand side of (\ref{2.1}) equated to zero. Such substitution yields $\Lambda$ independent of $x^{\mu}$ and $\partial_{\nu}
\tilde{\Theta}_{\mu}(x)+\partial_{\mu}\tilde{\Theta}_{\nu}(x)=0$, whose solution is $\tilde{\Theta}_{\mu}(x)=\omega_{\mu\nu}x^{\nu}+ \Theta_{\mu}$ and $\,\omega_{\mu\nu}= -\omega_{\nu\mu}$. These correspond to the ten Killing vectors (6 for $\omega_{\mu\nu}x^{\nu}$ and 4 for $\Theta_{\mu}$) under which Minkowski spacetime metric is invariant. They are associated with Lorentz transformations and spacetime translations, respectively \cite{wein2}. As we will see below $\Theta_{\mu}$ can be identified with the four velocity of the fluid as a whole and thus it inherits spacetime symmetries in the form of rigid motions: world lines of neighbor fluid elements would keep their separation whenever they lie along Killing vectors \cite{Israel1963}. We shall restrict such motions to translations in the present work and so only $\Theta_{\mu}$ is considered.

To determine  $\alpha$ and $\Theta^{\mu}$  one assumes that typical physical quantities are related to the statistical moments of the distribution. For instance the particle number density flux and the energy-momentum tensor
\begin{eqnarray}
N^{\mu } &=& c \int \frac{d^3p}{p^0} p^{\mu} \bar{f}\;, \label{Nflux}\\
T^{\mu \nu} &=& c \int p^{\mu}p^{\nu} \bar{f} \frac{d^dp}{p^0}\;,\label{2.39}
\end{eqnarray}
can be respectively written as
\begin{eqnarray}
    N^{\mu} &=& -\alpha c \frac{\partial {\cal I}}{\partial \Theta_{\mu}}\;,\label{NderTheta}\\
    T^{\mu \nu} &=& \alpha c \frac{\partial^2 \mathcal{I}}{\partial \Theta_{\mu}\partial \Theta_{\nu}}\;, \label{2.40} \\
    \mathcal{I} &:=& \int \frac{d^3p}{p^0} e^{ -\Theta_{\alpha}p^{\alpha} }\,, \label{I}
\end{eqnarray}
from which the denomination of $\cal I$ as generating functional suggests itself \cite{Israel1963}. Having explicitly defined $\cal I$ will allow to determine $\alpha$. On the other hand
to obtain $\Theta^{\mu}$ one relates the kinetic theory form for thermodynamical quantities with equilibrium thermodynamic equations assumed to hold in a frame comoving with the gas as a whole. This will relate $\Theta_{\mu}$ with a comoving temperature.

To evaluate (\ref{I}) one can express it in spherical coordinates on the mass shell in momentum space \cite{sygrg,Israel1963,neuge}, namely, since $p^{0}=\sqrt{\mathbf{p}^2+m^2c^2}$, one has for the components of the particle's momentum
 \begin{eqnarray}
 p^{\mu} &=& (mc\cosh\chi,\; mc\sinh{\chi}\sin\theta\cos\phi,\;\nonumber\\
 && mc\sinh\chi\sin\theta\sin\phi, \; mc\sinh\chi\cos\theta)\;, \\
 && 0\leq \chi\leq\infty,\; 0\leq \theta\leq \pi,\; 0\leq\phi\leq 2\pi\;, \nonumber
\end{eqnarray}
on the pseudosphere $p^{\mu}p_{\mu}=m^2c^2$. Hence (\ref{I}) becomes
\begin{eqnarray}\label{Ipseudo}
{\cal I} &=& \int d\Omega^{(3)} d\chi \sinh^2\chi\; \mathrm{e}^{-mc\Theta\cosh\chi}\nonumber\\
&=& 4\pi\frac{K_1(mc\Theta)}{mc\Theta}\;,\quad \Theta^{\mu}\Theta_{\mu}=\Theta^2\;,
\end{eqnarray}
where $d\Omega^{(3)}$ is the element of solid angle in three spatial dimensions, $K_1$ is the modified Bessel function or order one \cite{abrwtz} and $\Theta^{\mu}$ has been chosen to lie along the $\chi=0$ axis. Upon use of the identity between Bessel functions $\frac{d\mbox{}}{du}[u^{-n}K_{n}(u)]=-u^{-n}K_{n+1}(u)$ to evaluate the components of the statistical moments (\ref{NderTheta}) and (\ref{2.40}) allows to relate their components in the comoving frame to get the equation of state for the gas ${\cal P}=\rho c^2 \Theta$, with ${\cal P}$ the pressure and $\rho$ the density of energy, both in the comoving frame. This led Israel \cite{Israel1963} to identify $\Theta=\frac{c}{kT}$, with $T$ the invariant comoving temperature.

A different way to deal with (\ref{I}) is to consider that, being invariant, it can be calculated in a convenient frame, say one in which $\Theta^{\mu}=(\Theta^0,\mathbf{\Theta}=\mathbf{0})$. Here cartesian coordinates have been used in momentum space  \cite{dG,K&C}.
Noticing that (i) $\Theta_{\mu}$ in (\ref{Nflux}) should be timelike to make the integral to converge and (ii) the only available timelike vector for the gas as a whole is its velocity ${\mathscr U}^{\mu}$, one is led to propose $\Theta^{\mu}=\frac{{\mathscr U}^{\mu}}{kT}$, so that $T$ is identified with a quantity in a frame in which $\boldsymbol{{\mathscr U}}$, the spatial part of ${\mathscr U}^{\mu}$, is zero. Such frame is the comoving frame to the gas. This approach in which one picks the comoving frame from scratch begs the question about which is the Lorentz transformation of the temperature.

It would be desirable to be able to combine the power of the spherical components of the manifestly covariant approach mentioned afore  with the more intuitive and easier to handle cartesian ones and still be able to investigate the behavior of temperature under Lorentz transformations. This is indeed possible and will be discussed in the next section.

\section{Manifestly covariant generating functional with cartesian coordinates}
 Since our analysis remains the same independently of the number of spatial dimensions we consider such general case at once. Hence unless otherwise stated, tensor indices run as follows: $\mu,\nu\dots= 0,1,2,\dots,d$, for $d$ spatial dimensions.

\subsection{Determination of $\alpha$}
 Let us consider a frame non-comoving with the gas so that the spatial components of $\Theta^{\mu}$, $\mathbf{\Theta}$, are non-zero.
In $d$-dimensional space the integral $\mathcal{I}$, Eq. (\ref{I}), just requires to change the measure from $d^3p$ to $d^dp$. Now we adopt spherical coordinates only for the spatial part of the momentum and assume that $\mathbf{\Theta}\cdot\mathbf{p} =|\mathbf{\Theta}| |\mathbf{p}|\cos\vartheta_1$. We then have $d^dp=|\mathbf{p}|^{d-1}d|\mathbf{p}|d\Omega^{(d)}$, where $d\Omega^{(d)} = (\sin \vartheta_1)^{d-2}(\sin \vartheta_2)^{d-3}\ldots(\sin \vartheta_{d-2})^1 d\vartheta_1d\vartheta_2\ldots d\vartheta_{d-2}d\varphi = \prod_{i=1}^{d-2}(\sin \vartheta_i)^{d-i-1}d\vartheta_id\varphi$, where $0<\varphi <2\pi$ y $0<\vartheta_i <\pi$ \cite{path}. Thus $\mathcal{I}$ can be written as
\begin{equation}\label{2.10}
      \mathcal{I} = S_{d-1}  \int  \frac{d|\mathbf{p}||\mathbf{p}|^{d-1}}{p^0}  (\sin{\vartheta_1})^{d-2}d\vartheta_1 e^{ -\Theta_0p^{0} } e^{|\mathbf{\Theta}|
      |\mathbf{p}|\cos{\vartheta_1}}.
\end{equation}
Here $ S_{d-1}=\frac{2\pi^{\frac{d-1}{2}}}{\Gamma\left(\frac{d-1}{2}\right)}$  is the hyper-surface of the $d-1$ dimensional sphere \cite{path} resulting from 	 integrating over $\varphi$ and $d\Omega^{(d-1)}$, which excludes $\vartheta_1$.
To integrate over $\vartheta_1$ it is better to use the series form of the spatial exponential $
    e^{|\mathbf{\Theta}||\mathbf{p}|\cos{\vartheta_1}} = \sum^{\infty}_{k=0}
    \frac{\left(|\mathbf{\Theta}||\mathbf{p}|\cos{\vartheta_1}\right)^k}{k!}
$. In this way (\ref{2.10}) becomes
 \begin{eqnarray}\label{2.12}
 \mathcal{I} &=& S_{d-1}  \sum^{\infty}_{k=0} \frac{|\mathbf{\Theta}|^k}{k!}  \int  \frac{|\mathbf{p}|^{d-1+k}d|\mathbf{p}|}{p^0} e^{ -\Theta_0p^{0} }\times\nonumber\\
&&       \times \int^{\pi}_0 \sin{\vartheta_1}^{d-2}{\cos{\vartheta_1}}^k d\vartheta_1.
\end{eqnarray}
The remaining angular integral is non-zero only for $k=2n, n=0,1,2,\dots$ and, moreover, it can be related to the beta functions $B\left(\frac{2n+1}{2},\frac{d-1}{2}\right)$ \cite{abrwtz} so that
\begin{eqnarray}\label{2.15}
\mathcal{I} &=& S_{d-1}  \sum^{\infty}_{n=0} \frac{|\mathbf{\Theta}|^{2n}}{(2n)!} B\left(\frac{2n+1}{2},\frac{d-1}{2}\right) \nonumber\\
&&      \times \int  e^{ -\Theta_0p^{0} }\left[ {p^{0}}^2 - m^2c^2
      \right]^{\frac{2n+d-2}{2}}dp^0.
\end{eqnarray}
Where we have exchanged the independent variable $|\mathbf{p}|$ by $p^0$ by using the mass shell condition. The further change $y^0 = p^0/mc$ and $z_0 =mc\Theta_0$ together with the integral form $\int_1^{\infty}dx \mathrm{e}^{-ax}(x^2-1)^{m-\frac{1}{2}} = \left(\frac{2}{a}\right)^{m}\frac{\Gamma(m+\frac{1}{2})}{\Gamma(\frac{1}{2})} K_{m}(a)$ for the modified Bessel function \cite{abrwtz} leads to
\begin{eqnarray}\label{2.17}
\mathcal{I} &=& S_{d-1}  \sum^{\infty}_{n=0} \frac{|\mathbf{\Theta}|^{2n}}{(2n)!} B\left(\frac{2n+1}{2},\frac{d-1}{2}\right) (mc)^{2n+d-1}\times\nonumber\\
&&      \times \left(\frac{2}{z_0}\right)^{\frac{2n+d-1}{2}}\frac{\Gamma\left(n+\frac{d}{2}
      \right)}{\Gamma\left(\frac{1}{2}\right)}K_{\frac{2n+d-1}{2}}(z_0).
\end{eqnarray}
At this point we make the following considerations: (i) We use the known expression of the beta function in terms of the gamma functions \cite{abrwtz} and then (ii) introduce $z^{\mu}= mc\Theta^{\mu}$ together with $\beta:=|\mathbf{z}|/z^0$, $0\leq \beta < 1$, the latter following from the fact that $\Theta^{\mu}$ is timelike. Thereby Eq. (\ref{2.17}) may be rewritten as

\begin{equation}\label{2.23}
      \mathcal{I} = 2^{\frac{d+1}{2}} (mc)^{d-1}  \left( \frac{\pi}{z_0} \right)^{\frac{d-1}{2}} \sum^{\infty}_{n=0} \beta^{2n}z_0^{n} \frac{K_{n+\frac{d-1}{2}}(z_0)}{2^n n!}.
\end{equation}

Here we arrive at a critical point from the technical perspective. The sum in (\ref{2.23}) can be further reduced to a modified Bessel function upon using the multiplication theorem of the Bessel functions \cite{watson}
\begin{equation}\label{2.25}
    K_{\nu}(\lambda x) =\lambda^{\nu} \sum_{m=0}^{\infty}
    \frac{(-1)^m(\lambda^2-1)^m(\frac{1}{2}z)^m}{m!}K_{\nu+m}(x),
\end{equation}
with $|\lambda^2-1 |<1$. This finally produces
\begin{equation}\label{2.29}
      \mathcal{I} = 2 (mc)^{d-1}  \left(2 \frac{\pi}{z} \right)^{\frac{d-1}{2}}  K_{\frac{d-1}{2}}\left(z\right).
\end{equation}
Formula (\ref{2.25}) is essential to trade the Bessel series in (\ref{2.23}) by a single modified Bessel function in (\ref{2.29}).
This becomes (\ref{Ipseudo}) when $d=3$.
We see that $\mathcal{I}$ is only function of the invariant $z$. Now we can obtain the relativistic particle number density flux from (\ref{NderTheta}) producing
\begin{equation}\label{2.34}
    N^{\mu} =  2m^d c^{d+1} \alpha \left(2 \frac{\pi}{z} \right)^{\frac{d-1}{2}} K_{\frac{d+1}{2}}\left(z\right) \frac{z^{\mu}}{z}.
\end{equation}	
Such equation can be solved for $\alpha$ giving
\begin{equation}\label{alpha}
\alpha:=\frac{N}{2c(mc)^d K_{\frac{d+1}{2}}\left(mc\Theta\right)} \left(\frac{mc\Theta}{2\pi} \right)^{\frac{d-1}{2}}\;.
\end{equation}
It is clear from (\ref{2.34}) that $z^{\mu}$ and $N^{\mu}$ point in the same direction and since $N^{\mu} = N \mathscr{U}^{\mu}/c$ we have that
\begin{equation}\label{2.381}
\Theta^{\mu} = \frac{\Theta}{c} \mathscr{U}^{\mu}.
\end{equation}
The physical significance of $\Theta$  within our approach is discussed in the next subsection.

\subsection{$\Theta^{\mu}$ and the invariant Temperature}
 Now we seek to identify $\Theta^{\mu}$ with a thermodynamical quantity. Let us take as our starting point the Gibbs form of the second law of thermodynamics for a closed system \cite{tol,neuge}, which we shall assume to hold in the comoving frame
\begin{equation}\label{2.57}
    \delta U = T \delta \mathcal{S} - \mathcal{P}\delta V.
\end{equation}
Clearly we must relate the relativistic kinetic expressions like the energy-momentum tensor and entropy flux with internal energy, $U$, entropy, $\mathcal S$, pressure $\cal P$ and volume $V$ appearing in (\ref{2.57}).
To begin with let us introduce the distribution function (\ref{2.3}) and Eq. (\ref{alpha}), in the expression for the energy-momentum tensor (\ref{2.40}). This leads to
\begin{equation}\label{2.401}
    T^{\mu\nu} = -\frac{N}{\Theta}\left( \eta^{\mu \nu} - \frac{K_{\frac{d+3}{2}}(z)}{K_{\frac{d+1}{2}}(z)}\frac{\Theta^{\mu} \Theta^{\nu}}{\Theta}mc
    \right).
\end{equation}
The comoving pressure can be obtained from the energy-momentum tensor in $d$-dimensions as
\begin{equation}\label{2.52}
    \mathcal{P} \equiv -\frac{1}{d}\Delta_{\mu \nu}T^{\mu \nu} = \frac{N}{\Theta},
\end{equation}
where the projector is $\Delta^{\mu \nu}\equiv \eta^{\mu \nu}-\mathscr{U}^{\mu}\mathscr{U}^{\nu}c^{-2}$. The corresponding $d$-dimensional entropy flow (\ref{S3})
can be conveniently reexpressed as
\begin{equation}\label{2.50}
     S^{\mu} = -k\left[\ln{\left(\alpha h^d \right)}N^{\mu} -  T^{\mu \nu} \Theta_{\nu}\right].
\end{equation}

It is worth stressing that the quantities $N^{\mu}$, $T^{\mu \nu}$ and $S^{\mu}$ are densities, and therefore do not depend on the size of the system. It is however rewarding to include the fact that the gas is inside a box and to do so we make use of the characteristic function $\chi_{_{\mathrm{Box}}}$ defined in the introduction. In particular integrating over $d$ spatial dimensions
\begin{eqnarray}
  \mathcal{N} &=& c^{-1}\int \chi_{_{\mathrm{Box}}} N^{\mu} d\sigma_{\mu}, \label{x2}\\
  \mathcal{S} &=& c^{-1}\int \chi_{_{\mathrm{Box}}} S^{\mu} d\sigma_{\mu}. \label{x4}\\
    \mathcal{G}^{\mu} &=& c^{-1}\int \chi_{_{\mathrm{Box}}} T^{\mu \nu} d\sigma_{\nu}, \label{x3}
\end{eqnarray}
In the previous expressions (\ref{x2})-(\ref{x3}), $d\sigma_{\mu}={n}_{\mu}d^dS$, with ${n}_{\mu}$ a unit normal to the spatial hypersurface, and therefore a timelike vector; in particular $d^dS=d^dx$ when ${n}^{\mu}=\delta^{\mu}_0$. Obviously $\mathcal{N}$ and $\mathcal{S}$ are Lorentz invariants while $\mathcal{G}^{\mu}$ is a vector. The latter is just the total relativistic momentum of system.

Combining (\ref{2.50}) and (\ref{x4}) yields
\begin{equation}\label{x7}
    \mathcal{S} = -k \left[ \mathcal{N}\ln{\left( \alpha h^d \right)} - \Theta_{\mu}\mathcal{G}^{\mu} \right],
\end{equation}
Now we can make contact with (\ref{2.57}) by going to the comoving frame. Consider (\ref{x7}) for which such process means, in light of (\ref{2.381}),  $\Theta_{\mu}{\cal G}^{\mu}\rightarrow \Theta_0{\cal G}^{0}=\Theta{\cal G}^{0}$.
The differential of the entropy,  (\ref{x7}), in thermodynamic space ($\Theta$ constant) becomes hereby
\begin{eqnarray}\label{x9}
    \delta \mathcal{S} &=& k \Theta \left[ \delta \mathcal{G}^0 -  \frac{\mathcal{N}\mathcal{P}}{c}\frac{\delta N}{N^2} \right]\,,\nonumber\\
    &=& \frac{k \Theta}{c} \left[ \delta (c\mathcal{G}^0) +  \mathcal{P}\delta V \right]\,,
\end{eqnarray}
where use has been made of (\ref{alpha}), (\ref{2.52}) and the relation $\delta V = -\frac{\mathcal{N}\delta N}{N^2}$.
Comparison of (\ref{2.57}) with (\ref{x9}) gives rise to the identification
\begin{equation}\label{2.58}
    \Theta = \frac{c}{kT}.
\end{equation}
with $T$ the comoving temperature by interpreting $c{\cal G}^0$ as the relativistic generalization of internal energy $U$; this clearly holds in the low speed regime in which $c{\cal G}^0\approx {\cal N}mc^2 + U_{\mathrm{non-rel}}$, with $U_{\mathrm{non-rel}}$ the usual non-relativistic internal energy of the gas.
Thus (\ref{alpha}), (\ref{2.381}) and (\ref{2.58}) complete our analysis of the manifestly covariant distribution function which is expressed in terms of invariant quantities, $N,\Theta,m,c,\Theta^{\mu}p_{\mu}$, in the fashion
\begin{equation}\label{2.60}
    f(p) = \frac{N}{2c(mc)^d K_{\frac{d+1}{2}}\left(mc\Theta\right)} \left(\frac{mc\Theta}{2\pi} \right)^{\frac{d-1}{2}}  \exp\left({-\Theta_{\mu}p^{\mu}}\right).
\end{equation}
Eq. (\ref{MJ}) is obtained from (\ref{2.60}) by considering $d=3$ and using the comoving frame to the gas. It should be stressed however that in deriving (\ref{2.60}) no assumption was made in regard to the Lorentz transformation of temperature. The latter came from the requirement to reproduce equilibrium thermodynamics in the comoving frame.

\section{Covariant equipartition theorem}
 Equipartition theorems are usually considered  to connect temperature with averages of say kinetic energy in the non-relativistic case and certain functions of momenta in the relativistic case \cite{tol-eq,lan2}. Now, as we have shown in the previous section a manifestly covariant approach to relativistic equilibrium distribution function unveils the convenience of using an invariant temperature which is the one associated with the comoving frame of the gas. Thus it is of interest to investigate the role an invariant temperature plays within a manifestly covariant equipartition theorem.

The non manifestly covariant but relativistic equipartition theorem seems to have been first considered by Tolman \cite{tol-eq} and later by others \cite{komar}. Its use as a criterion to determine the Lorentz transformation of temperature was criticized by Landsberg \cite{lan2} who stressed that it can accommodate both an invariant as well as a transforming temperature under a change of frame. Recently Cubero \textit{et al}.  performed numerical simulations indicating the existence of an invariant temperature \cite{cub} on the basis of the relativistic equipartition theorem of \cite{lan2} thus hinting at the invariant temperature included in Landsberg's analysis. In this section we shall provide a manifestly covariant form of the equipartition theorem which not only  contains an invariant temperature but includes Landsberg's version. Moreover our form of the theorem is neatly expressed in terms of the total momentum of the gas.

Let us start by reexpressing the generating functional $\mathcal{I}$ (\ref{I}) in $(d+1)$-dimensional momentum space \cite{dG}
\begin{equation}\label{eq11}
    \mathcal{I} = 2\int e^{-\Theta_{\mu}p^{\mu}}\,\delta\left( p_{\sigma}p^{\sigma} - m^2c^2 \right)H(p^0)\,d^{d+1}p,
\end{equation}
where $H(p^0)$ is the Heaviside function that restricts (\ref{eq11}) to positive energies. The mass shell condition is accounted for by the Dirac's delta.

By covariantly extending the argument in \cite{tol-eq}, we next integrate by parts $d+1$ times -one for each $p^{\mu}$- and discarding the boundary terms at infinity we obtain
\begin{eqnarray}\label{eq12}
\mathcal{I} =
    -\frac{2}{d+1}\int p^{\nu} \frac{\partial}{\partial p^{\nu}} \left[ e^{-\Theta_{\mu}p^{\mu}}\,\delta\left( p_{\sigma}p^{\sigma} - m^2c^2 \right)  \right] H(p^0)\,d^{d+1}p\;.
\end{eqnarray}
To integrate over $p^0$ we apply the properties of the Dirac delta which finally yields
\begin{equation}\label{eq14}
    \mathcal{I} = \frac{1}{d}\int\, e^{-\Theta_{\mu}p^{\mu}} \left[ \left( \frac{|\mathbf{p}|}{p^0}\right)^2 + \Theta_{\mu} \frac{\partial p^{\mu}}{\partial p^i}p^i \right] \frac{d^{d}p}{p^0}.
\end{equation}
Remarkably, notwithstanding the second term in (\ref{eq14}) containing the spatial components $p_i$ it is Lorentz invariant overall.

To connect with the physical framework Eq. (\ref{eq14}) is plugged back into (\ref{NderTheta}) and (\ref{2.40}) to yield
\begin{eqnarray}\label{eq15}
    N^{\alpha} &=& \frac{c}{d} \int\, \frac{d^{d}p}{p^0} f \left\{ p^{\alpha} \left[ \left(\frac{|\mathbf{p}|}{p^0}\right)^2 +  p^i\frac{\partial p^{\mu}}{\partial p^i}\Theta_{\mu} \right] -  p^i \frac{\partial p^{\alpha}}{\partial p^i} \right\},
\\
\label{eq16}
    T^{\alpha \beta} &=& \frac{c}{d} \int\, \frac{d^{d}p}{p^0} f \left\{ p^{\alpha} p^{\beta} \left[ \left(\frac{|\mathbf{p}|}{p^0}\right)^2 +  p^i\frac{\partial p^{\mu}}{\partial p^i}\Theta_{\mu} \right] -  p^i \frac{\partial p^{\alpha}p^{\beta}}{\partial p^i} \right\}.
\end{eqnarray}
A further spatial integration of (\ref{eq15}) and (\ref{eq16}) according to (\ref{x2}) and (\ref{x3}) gives rise to
\begin{eqnarray}\label{eq17}
    d &=&\Theta_{\mu} \left\langle\left\langle p^i\frac{\partial p^{\mu}}{\partial p^i}  \right\rangle \right\rangle\\
\label{eq19}
    \mathcal{G}^{\alpha} &=& \frac{\mathcal{N}}{d}\left[ \Theta_{\mu} \left\langle\left\langle   p^i\frac{\partial p^{\mu}}{\partial p^i}  \,p^{\alpha}\right\rangle \right\rangle  - \left\langle\left\langle p^i \frac{\partial p^{\alpha}}{\partial p^i}\right\rangle \right\rangle \right].
\end{eqnarray}
where $\langle\langle\cdot\rangle\rangle:=\frac{1}{\cal N}\int \cdot\; d^{d}\mu$ has been used. It is worth stressing that Eq. (\ref{eq17}) is just the manifestly covariant form of the equipartition theorem corresponding to that advanced by Tolman and Landsberg \cite{tol-eq,lan2}, respectively, expressed using the invariant comoving temperature $T=c/k\Theta$, Eq. (\ref{2.58}). A comment regarding covariance of (\ref{eq17}) is here in order. Although it contains the spatial sum $p^i\frac{\partial\mbox{}}{{\partial p^i}}$, we arrive to it from the manifestly covariant equation (\ref{eq11}). This is analogous to recognizing the invariance of $\frac{d^3\mathbf{p}}{p^0}$. Clearly the argument holds also for both terms in the r.h.s. of Eq. (\ref{eq19}).

Observing that (\ref{eq17}) and (\ref{eq19}) contain a common term suggests that the relativistic momentum can be made to enter the covariant equipartition theorem. To proceed further we need to determine the first term in (\ref{eq19}).  First we combine (\ref{x3}) and (\ref{2.401}) to obtain the explicit form for the momentum
\begin{equation}\label{eq21}
    \mathcal{G}^{\alpha} = \mathcal{N} \frac{\Theta^{\alpha}}{\Theta^2} \left[ mc\Theta \frac{K_{\frac{d+3}{2}}\left(mc\Theta \right)}{K_{\frac{d+1}{2}}\left(mc\Theta \right)} - 1 \right].
\end{equation}
By equating the projections $\Theta_{\alpha}{\cal G}^{\alpha}$ of  (\ref{eq19}) and (\ref{eq21}) we identify
\begin{equation}\label{eq23}
    \frac{\Theta_{\alpha} \Theta_{\mu}}{d}  \left\langle\left\langle p^i\frac{\partial p^{\mu}}{\partial p^i}  \,p^{\alpha}\right\rangle \right\rangle  = \frac{K_{\frac{d+3}{2}}\left(mc\Theta \right)}{K_{\frac{d+1}{2}}\left(mc\Theta \right)}mc\Theta.
\end{equation}
From here we finally arrive at
\begin{eqnarray}\label{eq24}
    \frac{\Theta_{\mu}\mathcal{G}^{\mu}}{\mathcal{N}} - z &=& \mathcal{F}_d(z),\; z=mc\Theta\\
\label{eq25}
    \mathcal{F}_d(z) &:=&  z \frac{K_{\frac{d+3}{2}}\left(z \right)}{K_{\frac{d+1}{2}}\left(z \right)} - 1 - z\;.
\end{eqnarray}
Amusingly just in the same way as the energy of a particle having momentum $p^{\mu}_{\mathrm{particle}}$ is determined by an observer having velocity $U^{\nu}_{\mathrm{obs}}$ is given by $E_{\mathrm{obs}}={p_{\nu}}_{\mathrm{particle}}U^{\nu}_{\mathrm{obs}}$ we interpret $\frac{1}{\Theta}\Theta_{\mu}\mathcal{G}^{\mu}$ as the energy of the gas determined by the comoving observer.

\begin{figure}[h!]
\centering
\includegraphics[width=5.3in]{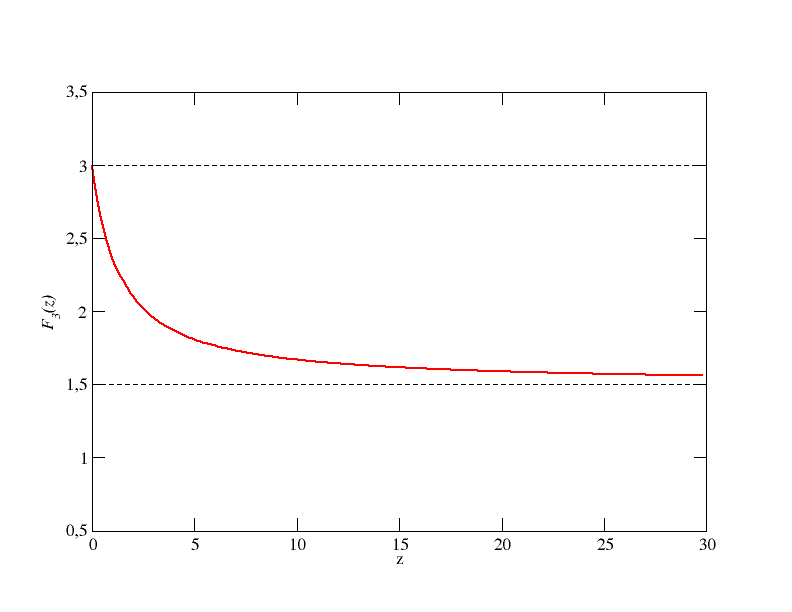}\\ 
\caption[Fvsz]{The graph shows $\mathcal{F}_d(z)$ vs. $z = mc^2/kT$ for $d=3$. $\mathcal{F}_3(z)$ corresponds to the quotient of the average of the relativistic kinetic energy per particle and the comoving temperature. We see that for $z\gg 1$ and $z\ll 1$ we get the appropriate limits, while for intermediate values of $z$, the ratio is a fraction between $3 < \mathcal{F}_3(z) < 3/2$.}
\label{f1}
\end{figure}

$\mathcal{F}_d(z)$ reduces to the usual values of the equipartition theorem in the limiting cases (See Fig. \ref{f1}).
\begin{equation}\label{eq26}
    \mathcal{F}_d(z) = \left\{
                         \begin{array}{ll}
                           \frac{d}{2}, & \hbox{$z\gg 1$,} \\
                           d, & \hbox{$z\ll 1$.}
                         \end{array}
                       \right.
\end{equation}

Notice that although a relativistic equipartition theorem had been considered before by other authors \cite{sygrg,haar} their approach was not manifestly covariant. Their work and ours coincide in terms of the function ${\cal F}_d$ (\ref{eq25}), in particular the behavior  in the non-relativistic as well as in the ultra-relativistic limits. Remarkably, recent numerical calculations adopting Monte Carlo methods \cite{peano} confirm ${\cal F}_d$ as giving the relativistic kinetic energy divided by $kT$.

\section{Discussion}

The interest in incorporating the relativity principles into kinetic theory goes beyond the theoretical foundations: Actual observations and experiments like for instance in high-energy physics \cite{rhic2}, astrophysics \cite{chandra} and cosmology \cite{wein} require a description of relativistic many-particles systems in terms of, say, Boltzmann's equation and the corresponding equilibrium distribution \cite{peack,bern,SZ}. Recently, Cubero \textit{et al}. \cite{cub} (See also references there) have developed numerical simulations based on Molecular Dynamics pointing to J\"{u}ttner's as the equilibrium distribution in agreement with their numerical analysis, as opposed to other proposals in the literature, for one and possibly other number of spatial dimensions. This was further confirmed for two \cite{montakhab,sim-eur} and three \cite{peano} spatial dimensions, the latter adopting Monte Carlo simulations instead. In \cite{cub}, the old problem regarding the relativistic transformation of temperature \cite{ei} was also considered in connection with a previous relativistic version of the Equipartition Theorem proposed by Landsberg \cite{lan2}. Based upon such theorem Cubero \textit{et al}. determined, within their approach, that temperature should possess an invariant character. It is worth mentioning that such analysis were not formulated in a manifestly covariant form and indeed some further insight was needed to interpret the contributions entering such theorem as well as temperature.

In this work we have obtained a new derivation of the manifestly invariant J\"{u}ttner's relativistic distribution function (\ref{2.60}). This is based on cartesian coordinates in ($d$+1)-momentum space ($d$ spatial dimensions) in contrast with the known results developed using spherical coordinates. This was made possible by the use of the multiplication theorem for Bessel's functions (\ref{2.25}) which simplified the treatment of a series involving Bessel functions \cite{watson}. In this approach no assumption is made a priori of any specific relativistic character for temperature. The latter appears through the invariant norm, $\Theta$, of a four-vector, $\Theta^{\mu}$ (\ref{2.381}), and it is invariant just for the same reason a point particle's rest mass is. Indeed because of the assumption that relativistic equilibrium kinetic theory should agree with standard thermodynamics for an observer comoving with the system under study the pseudo-norm becomes $\Theta=c/kT$, Eq. (\ref{2.58}), $T$ being the comoving temperature of the gas. Finally we have developed a manifestly covariant Equipartition Theorem, Eq. (\ref{eq24}), in which the average of the energy-momentum of the gas as determined by the comoving observer is given by a function ${\cal F}_d$, Eq. (\ref{eq25}) of the invariant temperature. Indeed this is analogous to the case of a point particle for which the energy is obtained by projecting momentum along the four-velocity of the observer. Here we have the thermal vector $\Theta^{\mu}=\frac{c}{kT}\mathscr{U}^{\mu}$, with $\mathscr{U^{\mu}}$ the four-velocity of the gas as whole which defines a comoving observer reading the invariant temperature $T$.
A further comment is here in order regarding the difference between our approach and previous ones. While previous versions of the Equipartition theorem \cite{lan2,tol-eq} relate temperature with a peculiar combination of relativistic quantities (See for example Eq.(\ref{eq17})), here we interpret that combination as included in (\ref{eq24}) which relates the invariant temperature with the averaged relativistic momentum, ${\cal G}^{\mu}$.

In a nutshell the manifestly covariant form of J\"{u}ttner's distribution leads naturally to consider the invariant comoving temperature to characterize the equilibrium regime. As an illustrative example of how this result applies to known cases let us consider the case of black body radiation. Let us recall the standard analysis of this problem \cite{MTW,peack,wein}: The Lorentz invariance of $\frac{I_{\nu}}{\nu^3},$ containing the specific intensity $I_{\nu}$ and the frequency $\nu$ of the photons,  implies the invariance of the Planckian distribution
\begin{equation}\label{Planck}
\frac{I_{\nu}}{2h\nu^3}=\frac{1}{\mathrm{e}^{\frac{hv}{kT}}-1}.
\end{equation}
Indeed for two frames in relative motion one should have
\begin{equation}\label{TR}
    \frac{h\nu}{kT} = \frac{h\nu'}{k\mathcal{T}'},
\end{equation}
where primed quantities refers to an observer that moves with respect to radiation and, in particular, $\mathcal{T}'$ is suggested as the non-comoving temperature. Since the photons suffer a Doppler shift:
\begin{equation}\label{TR-0}
    \frac{\nu}{\nu'} = \gamma (1-\beta \cos{\delta}),
\end{equation}
where $\delta$ is the angle between the momentum of the photon and the gas velocity whereas $\beta$ is the ratio between the gas speed and $c$. The quantity $\mathcal{T}'$ takes the anisotropic form
\begin{equation}\label{CMBT}
    \mathcal{T}' = \frac{T}{\gamma (1-\beta \cos{\delta})}.
\end{equation}	
Notice that we could alternatively write a manifestly invariant form for (\ref{Planck}) as $(\mathrm{e}^{\Theta_{\mu}p^{\mu}}-1)^{-1}$. In this way one can focus in the invariant product $\Theta_{\mu}p^{\mu}$. By evaluating it in the two frames mentioned above and using that for photons $p'^0=|\mathbf{p}'|$
\begin{equation}\label{TR0}
    \frac{h\nu}{kT} = \Theta^{'0}p^{'0}\left( 1-\beta \cos{\delta} \right).
\end{equation}
Considering the invariant comoving temperature $T$ the components $\Theta'^{\mu}$ become $\frac{\gamma }{kT}(c,\mathscr{U})$ and thus (\ref{TR0}) reduces to (\ref{TR-0}). This shows the consistency of adopting such invariant comoving temperature, trough $\Theta^{\mu}$, in the description of the black body radiation without resorting to $\mathcal{T}'$, Eq. (\ref{CMBT}). The latter can be regarded only as an auxiliary quantity for the following reasons. Firstly, actual observations of the Cosmic Microwave Background Radiation (CMBR) \cite{smoot} reveal there is a frame in which it presents the black body structure. Measurements however involve brightness depending on frequency rather than a non-comoving temperature (\ref{CMBT}). Secondly, in the case of massive particles one finds and obstacle to handle (\ref{CMBT}) (See for instance \cite{alf}). For massive particles we would have
\begin{eqnarray}
{\cal T}' &=& \frac{T}{\gamma(1-\beta_{p}\beta\cos\delta)}\\
\beta _{p} &=& \frac{|\mathbf{p}|}{p^0}\,, \nonumber
\end{eqnarray}
which makes no sense due to the dependence on the particles's momentum. However adopting an invariant comoving temperature is viable just for the same reason that the case for photons described above works.

Our work then adds to recent claims elaborated on the basis of an Unruh-DeWitt detector \cite{cm} pointing to the impossibility to have a relativistic transformation of temperature \cite{lan4}. In any case all these results reinforce the idea that temperature makes undisputable sense in the comoving frame.

There are several possible extensions of the present work which could be of interest. One possibility is to extend the analysis in the present work to the case of relativistic Fokker-Planck equation, say along the lines of \cite{GK}.
In other direction, we have seen that  in order to solve Boltzmann's equation for the equilibrium distribution the concept of rigid motions appeared. They can be seen related to the symmetries of Minkowski spacetime that we have here considered and specifically to the existence of Killing vectors. Although symmetric curved spacetimes have been studied previously for instance in \cite{stew,bern} there seems to be no attempt to incorporate the analogue of Noether's theorem in regard to the kinetic theory in particular noticing that the latter can be linked to relativistic quantum field theory \cite{dG,debbsch}. This would be particularly useful in the manifestly covariant approach. Generally covariant approaches to statistical mechanics have also been studied to investigate many particle systems in general relativistic theories \cite{rovmer}. It is a challenge to incorporate similar ideas to kinetic theory. Furthermore the very notion of spacetime has been considered in these terms \cite{Hu}.

\section*{ACKNOWLEDGEMENTS}

This work was partially supported by Mexico's National Council of Science and Technology, under grants CONACyT-SEP 51132F and a CONACyT sabbatical grant to HAMT. GCA wishes also to acknowledge support from CONACyT grant 131138. Enlightening comments from L. S. Garc\'{\i}a-Col\'{\i}n and D. Cubero are highly appreciated.

\end{document}